\documentclass[%
twocolumn, 
notitlepage, 
aps,
prd, 
longbibliography,
10pt, 
superscriptaddress,
floatfix,
letterpaper,
reprint,
bibnotes,
amsmath,amsfonts,amssymb,
]{revtex4-2}

\usepackage[hidelinks]{hyperref}
\hypersetup{colorlinks,
citecolor=blue,
urlcolor=blue
}

\usepackage{float}
\usepackage{graphicx}
\usepackage{dcolumn}
\usepackage{bm}
\usepackage[compat=1.1.0]{tikz-feynman}

\definecolor{linkcolor}{HTML}{399B03}
\definecolor{urlcolor}{HTML}{399B03}
\hypersetup{pdfstartview=FitH, linkcolor=linkcolor,urlcolor=urlcolor,colorlinks=true}
\hypersetup{
    unicode=false,          
    pdftoolbar=true,        
    pdfmenubar=true,        
    pdffitwindow=false,     
    pdfstartview={FitH},    
    pdfauthor={},         
    colorlinks=true,        
    linkcolor=blue,         
    citecolor=red,          
}

\begin{document}

\title{Quantum algorithm for imaginary-time Green's functions}

\author{Diksha Dhawan}%
\affiliation{%
 Department of Chemistry, University of Michigan, Ann Arbor, Michigan 48109, USA
}%
\affiliation{%
 Department of Chemistry,  Virginia Tech, Blacksburg, VA 24060, USA
}%
\email{ddhawan@vt.edu}
\author{Dominika Zgid}%
\affiliation{%
 Department of Chemistry, University of Michigan, Ann Arbor, Michigan 48109, USA
}%
\affiliation{%
 Department of Physics, University of Michigan, Ann Arbor, Michigan 48109, USA
}%
\email{zgid@umich.edu}

\author{Mario Motta}%
\affiliation{%
IBM Quantum, Almaden Research Center, San Jose, California 95120, USA
}%
\email{Mario.Motta@ibm.com}

\begin{abstract}
Green's function methods lead to ab initio, systematically improvable simulations of molecules and materials while providing access to multiple experimentally observable properties such as the density of states and the spectral function. The calculation of the exact one-particle Green's function remains a significant challenge for classical computers and was attempted only on very small systems. Here, we present a hybrid quantum-classical algorithm to calculate the imaginary-time one-particle Green's function. The proposed algorithm combines variational quantum eigensolver and quantum subspace expansion to calculate Green's function in Lehmann's representation. We demonstrate the validity of this algorithm by simulating H$_{2}$ and H$_{4}$ on quantum simulators and on IBM's quantum devices.
\end{abstract}

\maketitle

\section{Introduction}

The solution of the time-independent Schr\"{o}dinger equation is one of the central challenges of computational many-body quantum mechanics. For the system of interest, such a solution provides access to multiple properties, such as excitation energies, ionization potentials, electron affinities, multipole moments, and optimized geometries. While traditionally in quantum chemistry, the wavefunction formalism is employed to obtain solutions of the Schr\"{o}dinger equation, the Green's function formalism provides an equally powerful theoretical and computational framework.

In quantum mechanics, Green's functions are defined as correlation functions, from which most commonly one extracts information about the system, such as the density of states, quasiparticle properties, and response functions. Moreover, the Green's function formalism also provides direct access to thermodynamic quantities such as Gibbs energy, entropy, or heat capacity while explicitly including the temperature dependence. Consequently, due to its direct access to spectra and thermodynamic quantities, the Green's function formalism is frequently employed \cite{ebert2000fully, yoffa1979electronic, yeh2022ab} in the study of solids.

Over the years, many approximate methods to compute Green's functions have been proposed, such as GW \cite{hedin1965new,aryasetiawan1998gw, Kotani1,Kotani2,stan2006fully, Yeh_GW_solids_2022}, the second-order Green's function method (GF2) \cite{holleboom1990comparison, dahlen2005self, phillips2014communication, Welden_gf2_thermo,Iskakov_gf2_bandgaps}, and Green's function coupled cluster (GFCC) \cite{nooijen1992coupled,kowalski2014coupled,zhu2019coupled,shee2019coupled, shee_gfccsd_envir,shee_gfccsd_solids,shee_gfccsdt}, and they have been applied to numerous molecular and condensed-matter problems. GF2, GW, and other methods based on diagrammatic expansions provide accurate results in the weakly and moderately correlated regimes.
However, many strongly correlated systems fall outside the regime of validity of these approximations. 

Embedding Green's function methods such as dynamical mean-field theory (DMFT) \cite{kotliar2004strongly,kotliar2006electronic,georges1996dynamical} and self-energy embedding theory(SEET) \cite{Zgid_2015,lan2015communication,lan2017generalized,rusakov2018self, Zgid_2017,Iskakov_NiO_SEET}, have been proposed to overcome this challenge. 
In these methods, a subset of strongly correlated orbitals is treated with a highly accurate method (referred to as a solver).  Such a solver is required to describe electronic correlation more accurately than other computationally less expensive, approximate methods such as GW or GF2.
The most common solvers are based on the full configuration interaction (FCI) ~\cite{caffarel1994exact,von1984computational} and its truncations ~\cite{Zgid_Chan_CI, Zgid_Chan_CI2}.
The exponential scaling of FCI severely limits the number of impurity orbitals that can be treated by them. The use of truncated methods, on the other hand, compromises with the accuracy of these solvers.

Recent developments in quantum computing have shown promise in overcoming these limitations. Several algorithms have been proposed for obtaining the Green's function using quantum machines. The majority of these algorithms focus on calculating the real-time Green's function. Real-time Green's functions provide us access to various experimental properties, including spectra, but they require the time evolution of a state.
Early works in this field include algorithms based on phase estimation~\cite{bauer2016hybrid, kreula2016non, kosugi2020construction} and quantum Lanczos recursion~\cite{baker2021lanczos}. Despite their accuracy and scalability, these fault-tolerant algorithms require longer coherence times, often exceeding the capabilities of contemporary quantum devices. In contrast, algorithms based on variational ansatz-based simulation of time evolution are noise-resilient~\cite{mcardle2019variational,yao2021adaptive,libbi2022effective}, and require fewer qubits and shallower circuits.
While promising, variational approaches face challenges, especially in the choice of an ansatz, which can compromise the accuracy of the results \cite{d2023challenges}. Additionally, we should consider the scalability of nonlinear parameter optimizations, which can possibly suffer from barren plateaus ~\cite{bittel2021training, cerezo2021variational}. 

In Ref~\cite{endo2020calculation,libbi2022effective}, the authors proposed the use of variational quantum simulation (VQS) to time-evolve the system. In an alternative approach, presented in Ref~\cite{endo2020calculation}, a Lehmann's representation \cite{fetter2012quantum} of Green's function is used, and the excited states are obtained through the subspace-search variational quantum eigensolver (SSVQE). In other works, SSVQE is replaced by the quantum-equation of motion (q-EOM) method~\cite{ollitrault2020quantum,rizzo2022one}. Jamet et al.~\cite{jamet2022quantum} calculated the continued-fraction representation of the Green's function in the Krylov subspace using quantum subspace expansion (QSE). In Ref.~\cite{keen2022hybrid,steckmann2023mapping}, the authors simplify the time-evolution unitary by using GFCC and Cartan's decomposition, respectively. Some recent works~\cite{keen2022hybrid,steckmann2023mapping,gomes2023computing} use McLachlan’s variational principle for Hamiltonian simulation. There are similar works that focus on response properties~\cite{kumar2023quantum,huang2022variational} or molecular spectra~\cite{colless2018computation, motta2023quantum}.

Some of these aforementioned approaches require time evolution of a wavefunction over a mesh of points. Consequently, the total simulation time has to be at least $T = 2\pi / \omega_{min}$, where $\omega_{min}$ is the lowest frequency to be resolved. Furthermore, the size of the mesh has to be at least $n = T/\Delta t > \Delta \omega / \omega_{\min}$, where $\Delta t$ is the spacing between adjacent points in the mesh, and $\Delta \omega$ is the desired frequency resolution.
Suppose that the mesh comprises of $\Delta \omega/\omega_{min}$ points, a quantity that can be in the order of tens or hundreds when $\Delta \omega$ and $\omega_{min}$ are of different orders of magnitude. In order to compute a function $f(t)$ for all points in the mesh, one needs to run $\Delta \omega/\omega_{min}$ circuits (one for each point in the mesh) and, for each such circuit, measure an operator whose expectation value is equal to $f(t)$. Resolving an expectation value within a target precision requires gathering a sufficiently high number of measurement outcomes (or shots), which in turn determines a considerable computational overhead. Furthermore, the implementation of time-evolution circuits is challenging on near-term quantum devices, since a single step of time evolution requires a circuit depth of the order 1000 for 10 spatial orbitals using low-rank decompositions~\cite{motta2021low} or qDRIFT~\cite{campbell2019random}.

Since our objective in calculating the Green's function is to access stationary state properties when the state is at equilibrium, we can avoid the high cost of time evolution by calculating the Green's function on the imaginary-time axis (instead of doing the time evolution on the real-time axis). 
Green's functions evaluated on the imaginary-time axis are smoother when compared to the ones evaluated on real-time axis, and converge to zero in the high-frequency limit. This smoothness makes finite-temperature self-consistent calculations on the imaginary-time axis numerically easier and more stable, while retaining the same equilibrium state information as provided by the real-frequency Green’s function~\cite{fetter2012quantum}.  Moreover, analytical continuation~\cite{klepfish1998analytic,JARRELL1996133,Vidberg1977,Nevanlinna,Caratheodory} can be, at least in principle, used to obtain real frequency Green's function from Matsubara
Green’s functions.

In this work, we explore the combined use of the variational quantum eigensolver (VQE) and the quantum subspace expansion (QSE) methods to calculate the Lehmann's representation of the imaginary-time, one-particle Green's function.

We avoid time evolution to calculate excited states and instead use the QSE algorithm to calculate the transition amplitudes necessary for expressing the Green's function in the Lehmann's representation. Our algorithm is a hybrid quantum-classical approach, since the approximation of Hamiltonian eigenpairs and the computation of transition amplitudes are performed on a quantum device, whereas the evaluation of the Lehmann's representation of Green's function is carried out using a classical device.
We intend to use this method in conjunction with an effective Hamiltonian approach called Dynamic Self-Energy Mapping (DSEM)~\cite{dhawan2021dynamical}. DSEM provides us with an effective sparse Hamiltonian, reducing the circuit depth compared to using a full molecular Hamiltonian of the system.
 
\section{Method}

\subsection{Green's Function Formulation}~\label{sec:GF_forulation}

The molecular electronic Hamiltonian in the Born-Oppenheimer approximation is given, in second quantization, by
\begin{equation}
 H = \sum_{ij}h_{ij}c_{i}^{\dagger}c_{j} + \frac{1}{2}\sum_{ijkl}v_{ijkl}c_{i}^{\dagger}c_{j}^{\dagger}c_{l}c_{k}
\;,
\end{equation}
where $h_{ij}$ are $v_{ijkl}$ are one-body and two-body integrals, respectively. Indices $i$, $j$, $k$, $l$ denote a set of molecular orthonormal spin-orbitals, and creation and annihilation operators are denoted as $c_{i}^{\dagger}$ and $c_{i}$. The exact one-body Green's function of such a system characterizes the propagation of a state containing an additional particle or hole and is given by
\begin{equation}\label{gf_exact}
    \iota G_{ij}(t, t^{'}) = \langle\Psi_{0}|T[c_{i}(t)c_{j}^{\dagger}(t^{'})]|\Psi_{0}\rangle
    \;,
\end{equation}
where $|{\Psi_{0}}\rangle$ is the ground state of the system, 
$\iota$ is the imaginary unit, $c_i(t) = e^{-\iota t H} c_i e^{\iota t H}$ is a field operator in the Heisenberg representation and $T$ stands for the time-ordered product of operators. In Eq.~\eqref{gf_exact}, time ordering only affects the convergence factor, and does not affect the Green's function value at any finite value of time. Therefore, we can rewrite Eq.~\eqref{gf_exact} as
\begin{equation}
    \iota G_{ij}(t, t^{'}) = \langle{\Psi_{0}}|\{c_{i}(t),c_{j}^{\dagger}(t^{'})\}|{\Psi_{0}}\rangle\Theta(t-t^{'})
    \;,
\end{equation}
where $\Theta(t-t^{'})$ is the Heaviside step function. The Matsubara Green's function, which is used to describe time-independent but temperature-dependent Green's functions, is then obtained by Fourier transforming Eq.~\eqref{gf_exact} to the frequency domain,
\begin{equation}
\begin{split}
G_{ij}(\iota \omega_{n}) &= \langle{\Psi_{0}}|c_{i}\frac{1}{\iota\omega_{n}+E_{0}-H}c_{j}^{\dagger}|{\Psi_{0}}\rangle \\
&+ \langle{\Psi_{0}}|c_{j}^{\dagger}\frac{1}{\iota\omega_{n}+H -E_{0}}c_{i}|{\Psi_{0}}\rangle,
\end{split}
\end{equation}
where $\omega_{n}$ are the Matsubara frequencies~\cite{boehnke2011orthogonal} defined as 
 \begin{equation}\label{matsubara_freq}
     \omega_{n} = \frac{(2n+1)\pi}{\beta}
 \end{equation}
 where $n$ is an integer and $\beta$ is the inverse temperature.
Subsequently, to arrive to the Lehmann representation form, we introduce the resolution of identity in the subspaces with $N-1$ and $N+1$ particles, where $N$ is the number of electrons in the ground state of the system, obtaining 
\begin{equation}
\label{lehmann}
\begin{split}
G_{ij}(\iota\omega_{n}) &= \sum_{\mu} \frac{ \langle{\Psi_{0}}|c_{i}|{\Phi_{\mu}^{+}}\rangle \langle{\Phi_{\mu}^{+}}|c_{j}^{\dagger} |{\Psi_{0}}\rangle }{\iota\omega_{n}+E_{0}-E_{\mu}^{+}} \\
&+ \sum_{\mu} \frac{ \langle{\Psi_{0}}|c_{j}^{\dagger}|{\Phi_{\mu}^{-}}\rangle \langle{\Phi_{\mu}^{-}}|c_{i}|{\Psi_{0}}\rangle }{\iota\omega_{n}+E_{\mu}^{-}-E_{0}}
\end{split}
\end{equation}
The first line of Eq.~\eqref{lehmann} represents a Green's function for the  electron attachment (EA) and the second line for the ionization potential (IP). 
The later contains all cationic states and produces poles in the IP spectrum. 
Analogously, EA Green's function results in poles associated with the anionic states. 
Now, let us define transition matrix elements as
\begin{equation}
\begin{split}
X_{\mu j}^{+} &= \langle{\Phi_{\mu}^{+}}|c_{j}^{\dagger}|{\Psi_{0}}\rangle; 
\;\;\; 
(X_{\mu j}^{+})^{*} = \langle{\Psi_{0}}|c_{j}|{\Phi_{\mu}^{+}} \rangle \\
X_{\mu i}^{-} &= \langle{\Phi_{\mu}^{-}}|c_{i}|{\Psi_{0}}\rangle; 
\;\;\; 
(X_{\mu i}^{-})^{*} = \langle{\Psi_{0}}|c_{i}^{\dagger}|{\Phi_{\mu}^{-}} \rangle.
\end{split}
\end{equation}
Substituting these matrix elements into Eq. \eqref{lehmann} gives us
\begin{equation}
\begin{split}
\label{lehmann_2}
G_{ij}(\iota\omega_{n}) &= \sum_{\mu} \frac{(X_{\mu i}^{+})^{*}X_{\mu j}^{+}}{\iota\omega_{n}+E_{0}-E_{\mu}^{+}}\\
&+ \sum_{\mu} \frac{(X_{\mu j}^{-})^{*}X_{\mu i}^{-}}{\iota\omega_{n}+E_{\mu}^{-}-E_{0}}.
\end{split}
\end{equation}
Using this form, we can calculate the Green's function if we can compute the ground- and excited-state energies along with the transition matrix elements. In this work, we employ VQE for ground-state calculations and QSE to compute excited-state energies and transition matrix elements.

\subsection{Quantum Subspace Expansion}~\label{sec:qse}
\begin{figure*}[t!]
\centering
\includegraphics[width=0.77\textwidth]{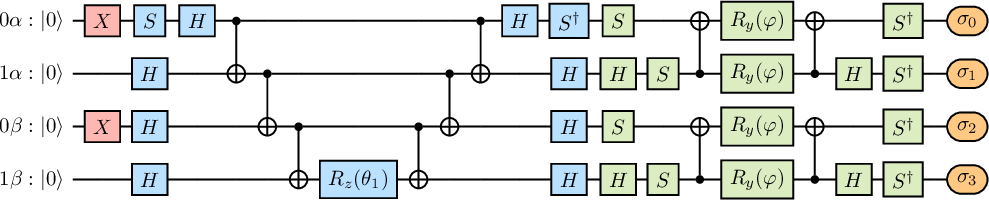}
\caption{Quantum circuit implementing a qubit coupled cluster ansatz for a two-electron in two-orbital system. Four qubits are used, storing the occupancies of the spin-up ($0\alpha,1\alpha$) and spin-down ($0\beta,1\beta$) orbitals in the standard Jordan-Wigner representation (qubits are ordered from top to bottom). The first group of Pauli $X$ gates (left, red) brings the qubits in the Hartree-Fock state $|0101\rangle$ (qubits are ordered from right to left). The second group of gates (center left,  blue) applies the exponential of the Pauli operator $\mathrm{XXXY}$, leaving the qubits in a state of the form 
$\cos(\theta_1/2) |0101\rangle + \sin(\theta_1/2) |1010\rangle$ (i.e. a linear combination of the Hartree-Fock state and a doubly-excited state with real-valued coefficients). The third group of gates (center right, green) implements a fixed unitary transformation rotating from the basis of molecular spin-orbitals to the basis of symmetrized atomic orbitals.
Finally, qubits are measured (right, orange ovals) to compute expectation values of Pauli operators ($P = \sigma_3 \sigma_2 \sigma_1 \sigma_0$) and, through classical post-processing, frequency-dependent Green's functions.
}
\label{fig:quantum_circuit}
\end{figure*}
QSE-based methods provide a compelling way of approximately solving the eigenvalue problem on a quantum computer \cite{bauer2020quantum,motta2021emerging}. The general idea of QSE is to obtain eigenspectrum of a system in a subspace defined by applying suitable excitation operators to a reference state, $|{\Psi}\rangle$. The ansatz for such a subspace can be defined as
\begin{equation}
    |{\Phi_i}\rangle = R_{i} |{\Psi}\rangle
    \;,
\end{equation}
where $|\Psi\rangle$ is the reference state, that can be obtained e.g. from a VQE calculation, and $\hat{R_{i}}$ is an excitation operator with the index $i$ specifying different levels of excitation. Then the eigenspectrum can be approximated by diagonalizing the Hamiltonian defined in the subspace spanned by $\{|{\Phi_{i}}\rangle\}$, essentially reducing the problem to a generalized eigenvalue problem,
\begin{equation}\label{eigval_eq}
    H^{sub}\, V = S^{sub} \, VE \; ,
\end{equation}
where $H^{sub}$ and $S^{sub}$ are the Hamiltonian and overlap matrices in the subspace of interest, $V$ are expansion coefficients defining approximate eigenstates as $| \Psi_\mu \rangle = \sum_i V_{i\mu} |{\Phi_{i}}\rangle$, and $E$ is a vector of approximate eigenvalues. The matrices $H^{sub}$ and $S^{sub}$ are defined as
\begin{equation}
    H_{ij}^{sub} = \langle{\Phi_{i}}|H|{\Phi_{j}}\rangle, \;\;\;
    S_{ij}^{sub} = \langle{\Phi_{i}}|{\Phi_{j}}\rangle
    \;.
\end{equation}
Here, the $H^{sub}$ and $S^{sub}$ matrices are computed on a quantum device and Eq.~\eqref{eigval_eq} is then solved on a classical computer to obtain eigenvectors and eigenenergies.
We note that the eigenstates $| \Psi_\mu \rangle$ are not prepared on a quantum computer, therefore computing excited-state properties and transition matrix elements generally requires measuring of additional matrices.

\subsection{Green's function calculation using QSE}

As discussed in Section~\ref{sec:GF_forulation}, we can use QSE to calculate the transition matrix elements. The Green's function can then be obtained from Eq.~\eqref{lehmann}. Here, for conciseness, we give a detailed description for the calculation of EA transition matrix elements only (the procedure is similar for IP transition matrix elements). Our implementation of QSE to calculate transition matrix elements is as follows:
\begin{itemize}
    \item Prepare the ground state wavefunction of the system. We note that this step is independent of the rest of the algorithm. We can use any quantum algorithm for the ground-state preparation. In this work, we use VQE.
    \item Choose an ansatz for the subspace with $N+1$ particles. In this work, we consider
\begin{equation}
    |{\Phi_i^{+}}\rangle = c_i^{\dagger} |{\Psi}\rangle
\end{equation}
where $c_i^{\dagger}$ is a creation operator. Here, we note that we are using only a subspace of the excited states, i.e. the linear response subspace. We choose to limit our subspace as a full space leads to exponential complexity~\cite{mcclean2017hybrid}. This choice leads to approximate eigenstates, that we assess in Sec.~\ref{sec_results}. 
 \item Calculate $H^{sub}$ and $S^{sub}$ matrices in the subspace with $N+1$ particles,
 \begin{equation}
    H_{ij}^{sub} = \langle\Psi|c_{i}Hc_{j}^{\dagger}|\Psi\rangle
    \;,\;
    S_{ij}^{sub} = \langle\Psi|c_{i}c_{j}^{\dagger}|\Psi\rangle
    \;,
\end{equation}
by measuring the expectation values of the operators $c_{i}Hc_{j}^{\dagger}$, $c_{i}c_{j}^{\dagger}$ over the wavefunction $\Psi$ on a quantum device.
\item Solve the eigenvalue equation
\begin{equation}
    H^{N+1}V = S^{N+1}VE.
\end{equation}
\item Compute transition matrix elements as
\begin{equation}
\begin{split}
X_{\mu j}^{+} &= \langle{\Phi_{\mu}^{+}}|c_{j}^{\dagger}|\Psi\rangle 
= 
\sum_{i} V_{i\mu}
\langle\Psi|c_{i}c_{j}^{\dagger}|\Psi\rangle \\
&= \sum_{i} V_{i\mu} S_{ij}
\;.
\end{split}
\end{equation}
\item Compute transition matrix elements in the subspace with $N-1$ particles similarly, i.e. by using annihilation operators in place of creation operators.
\item Substitute the values of the  transition matrix elements in Eq.~\eqref{lehmann_2} to compute the Green's function on classical computer.
\end{itemize}

\subsection{Statistical Analysis}
\label{sec:jackknife}

Statistically distributed data is an intrinsic part of any calculation performed on a quantum computer due to the randomness of the measurement process. We label the uncertainty in the measurements as ``shot noise''. Due to the shot noise, expectation values computed from quantum measurements are accompanied by statistical uncertainties, and the propagation of such statistical uncertainties is an essential part of any quantum calculation.

In particular, the matrix elements of $H^{sub}$ and $S^{sub}$ described in Sec.~\ref{sec:qse} are normally distributed. The standard deviations present in this data can be reduced by increasing the number of measurements (as the magnitude of sample variances is inversely proportional to the square of the number of measurements, also termed ``shots''), but remain finite and non-negligible.

When normally distributed data is subjected to non-linear transformations, such as Eq.~\eqref{lehmann} and Eq.~\eqref{lehmann_2}, the resulting distribution may become skewed and may not be symmetrically centered around the mean,
which makes reliable estimates of statistical uncertainties a delicate operation.
A natural and effective way to deal with such non-normally distributed data is using resampling techniques~\cite{efron1982jackknife} such as jackknife or bootstrap. In this work, we use jackknife~\cite{friedl2002jackknife,troyer2009quantum}. The jackknife method works by systematically redistributing the samples to form sub-samples, and estimating the statistics of each of these sub-samples separately. The general procedure for jackknife is as follows:
\begin{itemize}
    \item Distribute the samples into $M$ bins.
    \item Form a sub-sample $M_{i}$ out of $M-1$ bins, where $M_{i}$ contains all the bins except the $i^{th}$ one.
    \item Evaluate the mean, $U_{i}$, in each of the $M$ bins and obtain an error estimate from the variance of these estimates.
    \item The jackknife mean of the dataset is
    \begin{equation}
        U = U_{0}- (M-1)(\overline{U} - U_{0}),
    \end{equation} where $U_{0}$ is the average of the complete dataset, and $\overline{U}$ is the average of all $U_{i}$.
    \item The jackknife standard deviation of the dataset is 
    \begin{equation}
        \Delta U = \sqrt{M-1} \left(\frac{1}{M}\sum_{i=1}^{M}(U_i)^2 - \overline{U}^2\right)^{1/2}.
    \end{equation}
\end{itemize}
In section \ref{statistics}, we illustrate the use of the jackknife analysis to improve the statistics of our results.

\subsection{Computational Details}
In this section, we discuss the implementation details of our algorithm. We obtain our molecular Hamiltonians from PySCF~\cite{sun2018pyscf,sun2020recent} in the STO-6G basis. To obtain 
a sparser representation of the Hamiltonian than in the basis of molecular orbitals, we localize orbitals using a symmetrized atomic orbital (SAO) procedure, which produces a basis of orthonormal orbitals.

For quantum calculations, we use Qiskit~\cite{Qiskit}. We map the molecular Hamiltonian to a qubit operator using the standard Jordan-Wigner transform~\cite{batista2001generalized}, and compute the ground-state wavefunction using the VQE algorithm with the qubit coupled-cluster (QCC) Ansatz~\cite{ryabinkin2018qubit}. QCC applies exponentials of suitably-chosen Pauli operators $P_1 \dots P_m$ to the Hartree-Fock state,
\begin{equation}
|\psi_{\mathrm{QCC}}(\boldsymbol{\theta})\rangle 
= 
e^{K}
e^{-i \frac{\theta_{m}}{2} P_m} \dots e^{-i \frac{\theta_{1}}{2} P_{1}}
|\psi_{\mathrm{HF}}\rangle 
\label{eq:QCC1}
\;,
\end{equation}
where 
\begin{equation}
|\psi_{\mathrm{HF}}\rangle = \bigotimes_i \mathrm{X}_i | 0 \rangle^{\otimes 2n}
\end{equation}
is a computational basis state (also called a bistring) obtained initializing $2n$ qubits (as many as the spin-orbitals in the underlying basis set) in the state $|0\rangle$ and exciting qubits corresponding to occupied molecular spin-orbitals (denoted by the index $i$) with a Pauli $X$ operator, and $e^{K}$ is an active-space orbital rotation \cite{moreno2023enhancing,motta2023bridging} bringing from the basis of molecular orbitals into the basis of SAO.
In the Jordan-Wigner representation, the operator $e^{K}$ can be implemented \cite{jiang2018quantum,cohn2021quantum} by a quantum circuit of Givens rotations acting on adjacent qubits on a device with linear connectivity, using e.g. the design introduced originally by Reck et al \cite{reck1994experimental}, or the more economical one proposed by Clements et al \cite{clements2016optimal}.
\begin{figure}[!htbp]
\centering
e\includegraphics[width=\columnwidth]{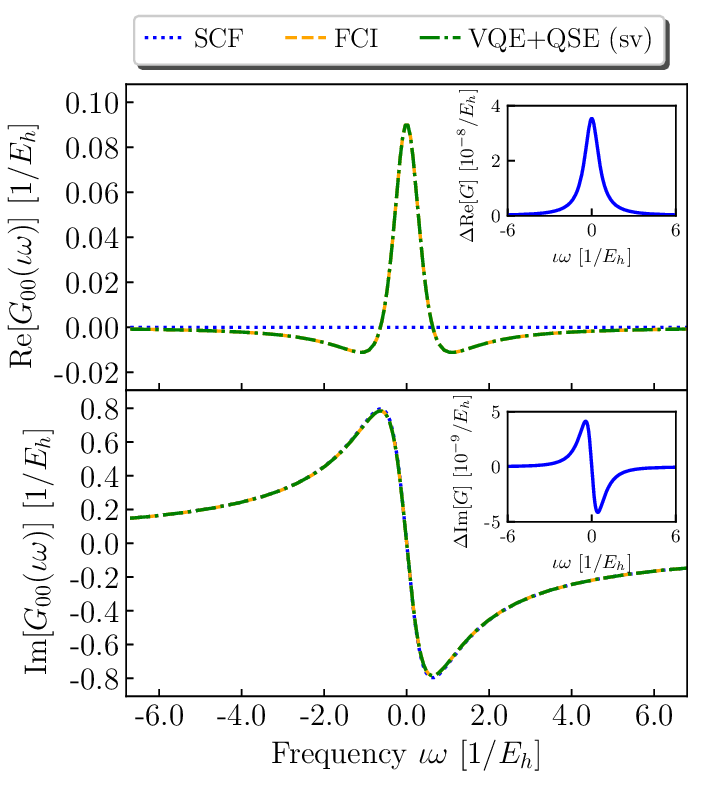}
\caption{Real (top panel) and imaginary part (bottom panel) of $G_{00}(\iota\omega)$ element for the H$_2$ molecule in the STO-6G basis evaluated using FCI and QSE. The inset  shows the difference between Green's functions calculated using FCI and QSE.}
\label{H2_GF00}
\end{figure}

\begin{figure}[!htbp]
\centering
\includegraphics[width=\columnwidth]{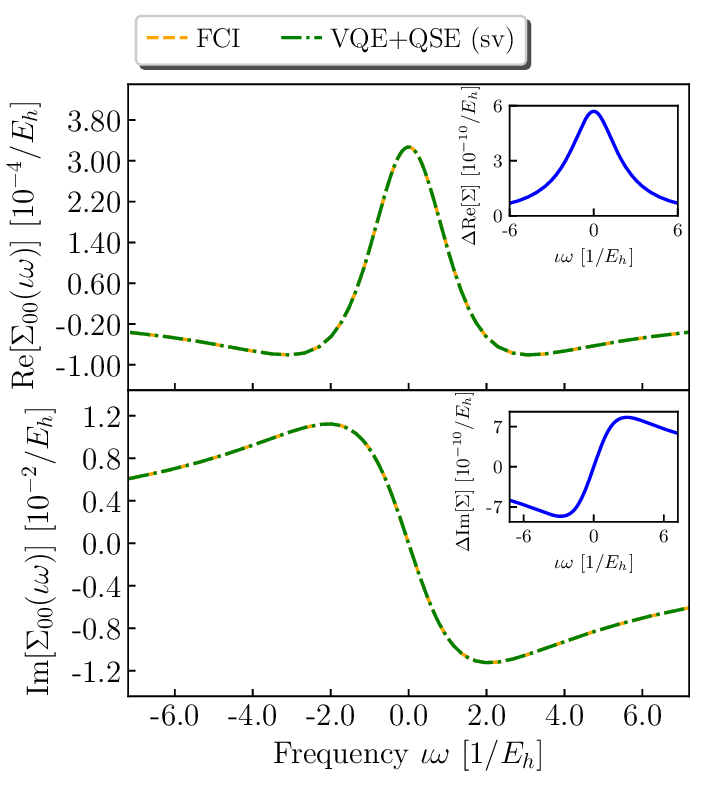}
\caption{Real (top panel) and imaginary part (bottom panel) of $\Sigma_{00}(\iota\omega)$ element for the H$_2$ molecule in the STO-6G basis evaluated using FCI and QSE. The inset  shows the difference between the self-energy elements calculated using FCI and QSE.}
\label{H2_SE00}
\end{figure}
In this work, we employed Pauli operators producing double and single electronic excitations on top of the Hartree-Fock state,
\begin{equation}
\begin{split}
P_{ijab} &= \mathrm{X}_b \mathrm{X}_a \mathrm{X}_j \mathrm{Y}_i \;, \\
P_{ia} &= \mathrm{X}_a \mathrm{Y}_i \;, \\
\end{split}
\end{equation}
where $i$, $j$ and $a$, $b$ label occupied and unoccupied spin-orbitals in the Hartree-Fock state, respectively, and $X$, $Y$ denote Pauli $X$ and $Y$ operators, respectively. We ordered these Pauli operators starting with those generating opposite-spin double excitations, followed by same-spin double excitations, and single excitations.
For two-orbital systems, with the aim of economizing hardware simulations, we employed a single Pauli operator in Eq.~\eqref{eq:QCC1}, specifically $P_1 = \mathrm{XXXY}$. The corresponding quantum circuit is shown in Fig.~\ref{fig:quantum_circuit}. While the quantum circuit is further optimized before the execution on a quantum hardware, we illustrate it before optimization for clarity.

We implement the VQE and QSE algorithms using Qiskit libraries. More specifically, we simulate quantum circuits using a classical noise-free
simulator, called $\mathsf{statevector}$ simulator, where exact unitary operations are applied to the state vector representing the wavefunction. This is used to assess the errors arising because of the approximations in the formalism developed in this work. 
\begin{figure*}[!ht]
\centering
\includegraphics[width=1.0\columnwidth]{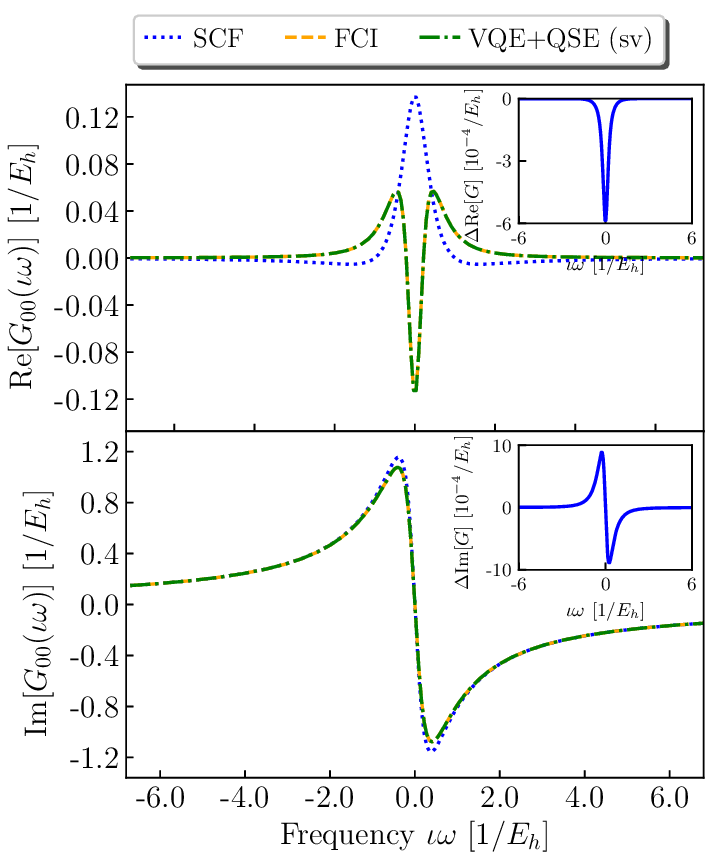}
\includegraphics[width=1.0\columnwidth]{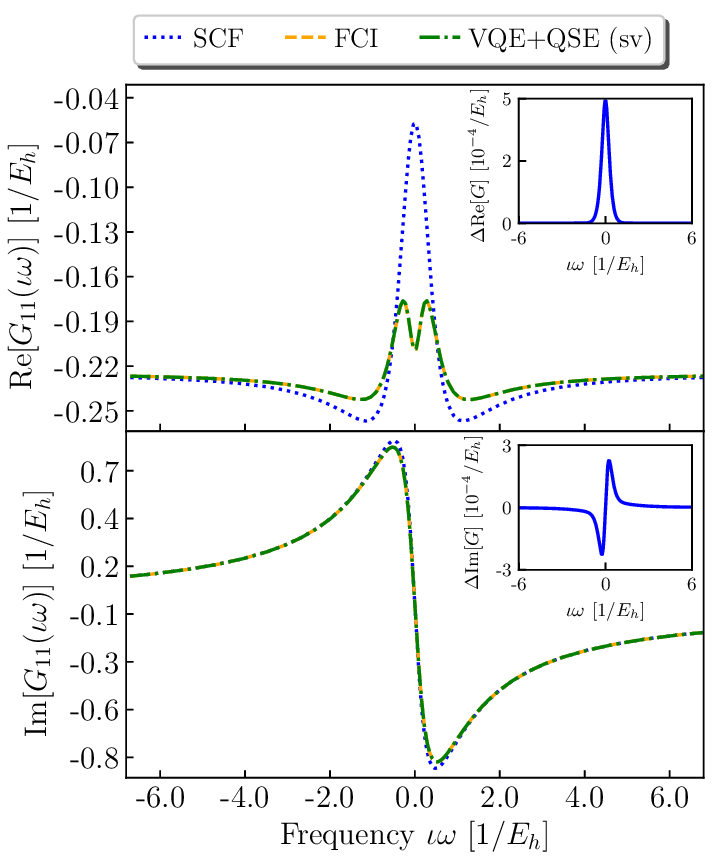}
\caption{Real (Left) and imaginary part (Right) of $G(\iota\omega)$ for the H$_4$ chain with bond length 1.0 {\AA} in the STO-6G basis evaluated using FCI and QSE. (a)$G_{00}(\iota\omega)$ element (Left) (b) $G_{11}(\iota\omega)$ element (Right) (The inset shows the difference between Green's functions calculated using FCI and QSE.)}
\label{H4_GF}
\end{figure*}

Separately, we assess the impact of the shot noise using the Quantum ASseMbly (QASM) simulator, which yields randomly distributed data (as opposed to the deterministic results from $\mathsf{statevector}$). Finally, we carry out our calculations on the IBM Quantum hardware, where we experience quantum decoherence along with approximation errors and the shot noise. To mitigate the impact of decoherence, we use the readout error mitigation \cite{nation2021scalable} and dynamical decoupling \cite{viola1998dynamical,kofman2001universal,biercuk2009optimized,rost2020simulation,niu2022effects,niu2022analyzing,ezzell2022dynamical} as implemented in the Qiskit Runtime library.
\section{Results}
\label{sec_results}

In this section, we present results that illustrate the method described above.
We test our method on H$_2$ and H$_4$ chains. The analysis of our results is structured in such a way as to differentiate between different sources of errors, whether coming from the approximation in the method, shot noise, and/or decoherence affecting quantum devices.

\subsection{Green's Function from QSE and FCI}
\label{statevector}

\begin{figure*}[!htbp]
\centering
\includegraphics[width=1.0\columnwidth]{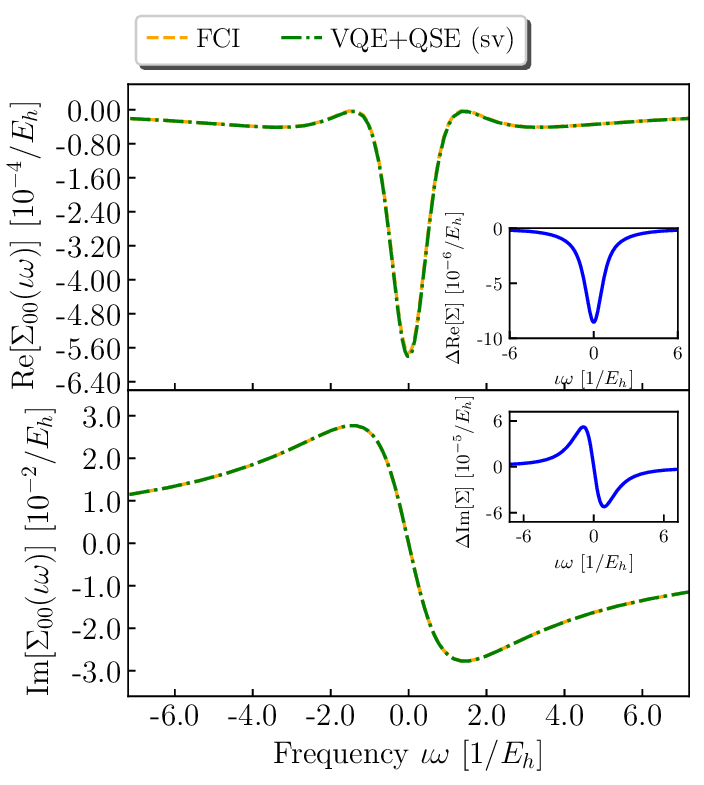}
\includegraphics[width=1.0\columnwidth]{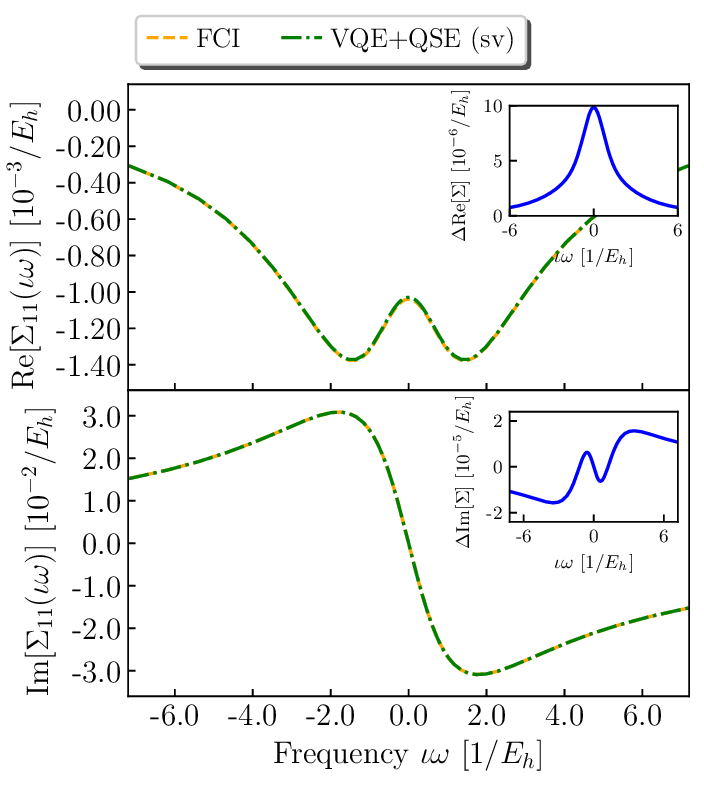}
\caption{Real (Left) and imaginary part (Right) of $\Sigma(\iota\omega)$ for the H$_4$ chain with bond length 1.0 {\AA} in the STO-6G basis evaluated using FCI and QSE. (a)$\Sigma_{00}(\iota\omega)$ element (Left) (b) $\Sigma_{11}(\iota\omega)$ element (Right)}
\label{H4_SE}
\end{figure*}

We begin our analysis of results by comparing the Green's functions obtained from QSE calculation performed on the $\mathsf{statevector}$ simulator to FCI Green's functions obtained on a classical machine. Since the results from the simulator are deterministic, this comparison serves to assess the errors arising from the approximations underlying the VQE and QSE algorithms. 
The real and imaginary part of $G_{00}(\iota\omega)$ for the H$_2$ molecule at equilibrium bond length, computed using QSE and FCI, are shown in Fig.~\ref{H2_GF00}. These Green's function matrix elements were computed using a Matsubara frequency grid~\cite{li2020sparse}. 
Since for this system, the real part of Green's function does not show large differences between HF and FCI, we further calculated the self-energy $\Sigma(\iota\omega)$ to highlight possible differences. 
The real and imaginary parts of $\Sigma_{00}(\iota\omega)$ are shown in Fig.~\ref{H2_SE00}. Additionally, we performed similar calculations on a H$_4$ chain with each hydrogen separated by 1 {\AA}. In Fig.~\ref{H4_GF}, we plot the $G_{00}(\iota\omega)$ and $G_{11}(\iota \omega)$ elements from both QSE and FCI. The inset in the plot shows the actual difference between the elements at different frequency points. Figures \ref{H2_GF00}-\ref{H4_GF} demonstrate that the results obtained from the QSE algorithm are in satisfactory agreement with FCI results.

\begin{figure*}[!t]
    \centering
    \includegraphics[width=2\columnwidth]{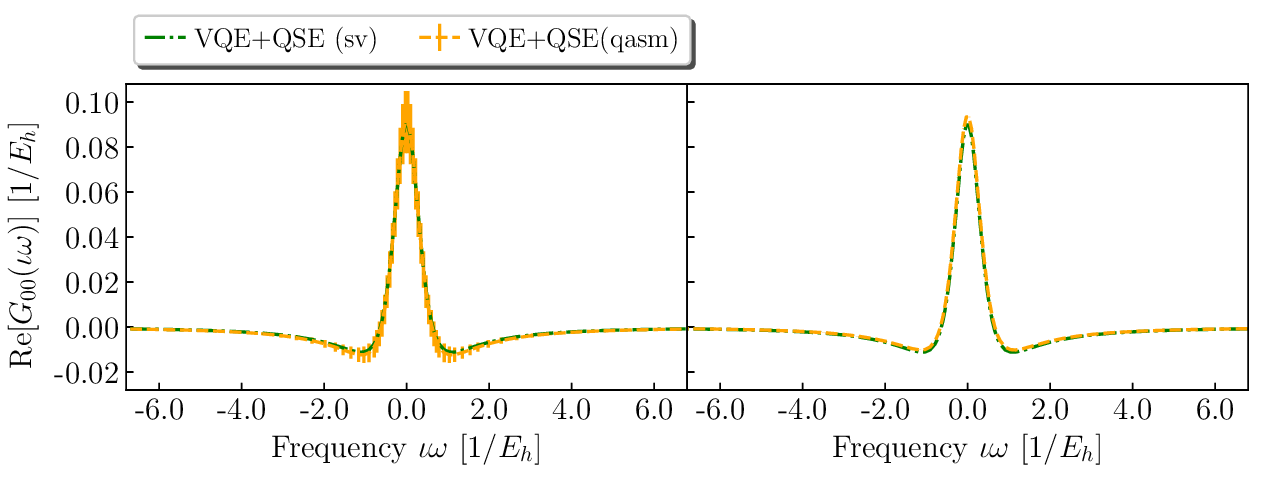}
\caption{The real part of $G_{00}(\iota\omega)$ element for the H$_2$ molecule with a bond length 0.76 {\AA} in the STO-6G basis evaluated using FCI and QSE done on the QASM simulator. The vertical bars represent the error-bars. (a) Without resampling (b) With resampling
}
\label{H2_qasm}
\end{figure*}
\subsection{Green's functions with measurement statistics}\label{statistics}
\begin{figure*}[!t]
\centering
\includegraphics[width=2\columnwidth]{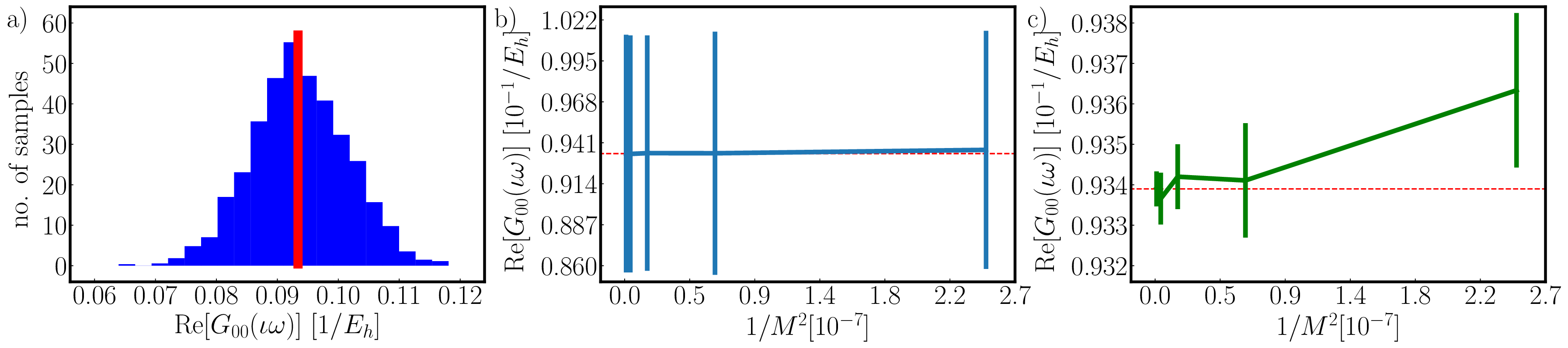}
\caption{a) Distribution of real component of the Green's function element $G_{00}(\iota\omega)$ for the samples taken through the QASM simulator. b) Error distribution obtained from the crude data (without using any resampling technique) as the sample size increases with respect to $1/M^2$ where M is the total number of samples. c) Error distribution obtained from resampled data as the sample size increases with respect to $1/M^2$ where M is the total number of samples.(Red line in b and c indicates mean value.)}
\label{H2_error}
\end{figure*}
Simulations on quantum devices result in statistically distributed data.
When Green's functions are evaluated using Eq.~\ref{lehmann} the normally distributed data are subjected to non-linear operations, as discussed in Sec.~\ref{sec:jackknife}.
\begin{figure}[!ht]
\centering
\includegraphics[width=\columnwidth]{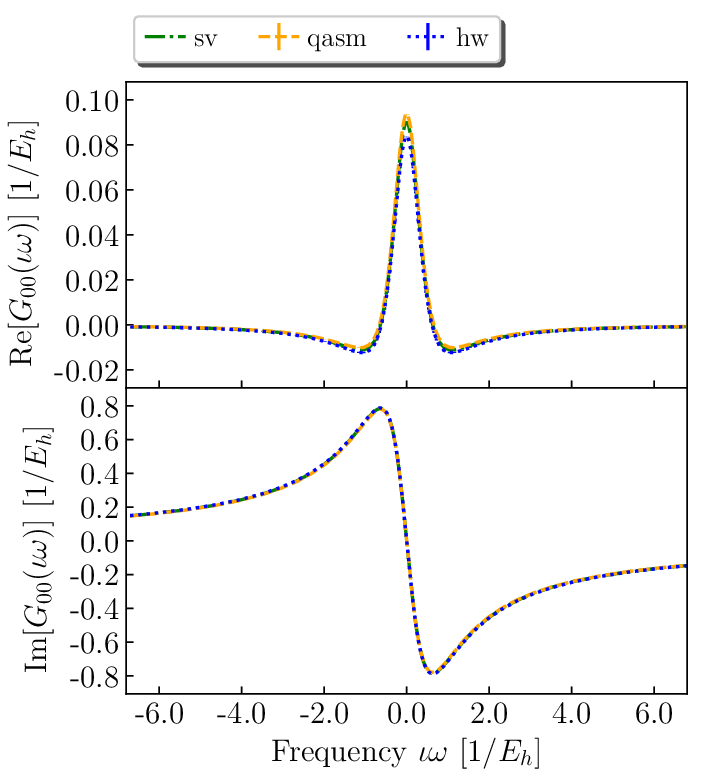}
\caption{Real (top) and imaginary (bottom) part of $G_{00}(\iota\omega)$ for the H$_2$ molecule with a bond length of 0.76 {\AA} in the STO-6G basis evaluated using QSE on $\mathsf{statevector}$ simulator, QASM simulator and hardware. The quantum hardware results have been post-processed using jackknife for a proper error propagation analysis. The vertical bars represent the error-bars.}
\label{H2_HW}
\end{figure}
In this section, we test how QSE behaves in the presence of shot noise and how statistical uncertainties propagate through non-linear operations necessary to estimate the final quantities, i.e. the elements of the Green's function. 
To achieve this goal, we perform calculations on the QASM simulator that mimics shot noise while avoiding any effects present due to decoherence and imperfect implementation of quantum operations on the quantum hardware.
        
The measurement data obtained from the simulator is expected to be normally distributed. However, as mentioned in the earlier section, non-linear operations on stochastic results render the distribution non-normal. For Green's function elements, this can be observed in Fig.~\ref{H2_error}a, illustrating the distribution of real and part of $G_{00}(\iota \omega)$ element for different number of statistical samples, where the histograms are deviating slightly from the normal behavior. The imaginary part of the element show similar behavior. This is further made clear by the behavior of the standard deviation with respect to number of samples, as shown in Fig.~\ref{H2_error}b. 

The standard deviation is expected to be inversely proportional to $M^2$, where $M$ is the number of samples. However, we observed that the standard deviation does not change significantly with an increasing number of samples. This observation, along with the histogram shown in Fig.~\ref{H2_error}a, suggests the presence of a fat-tail distribution~\cite{laherrere1998stretched}. A fat-tail distribution is characterized by more frequent occurrences of samples that are three or more standard deviations away from the mean value, compared to a normal distribution. To address the challenges posed by working with fat-tailed non-normal distributions, we employed the jackknife re-sampling technique. As shown in Fig. \ref{H2_error}c, this approach resulted in the expected decrease in error as the sample size increased.

In Fig. \ref{H2_qasm}, we compare the QSE Green's function calculated on $\mathsf{statevector}$ simulator to the Green's function calculated on QASM simulator after the jackknife. Note that QASM simulator simulates the measurement process but not the noise of near term devices. For further comparison, in Fig. \ref{H2_qasm}a, we show the Green's function element evaluated without the jackknife resampling technique. It is evident that the absence of resampling leads to a significantly larger error compared to the Green's function elements shown in Fig.~\ref{H2_qasm}b, that includes resampling.
\begin{table}[!ht]]
            \centering
            \begin{tabular}{@{}lccc@{}}
                \hline\hline
                 Figure & no. of electrons & \multicolumn{2}{c}{Max Error}\\ 
                \hline
                
                & & $Re[G_{00}]$ & $Im[G_{00}]$ \\
                \hline
                    Fig. \ref{H2_HW} & 1.99364 & 0.00637 & 0.00466\\
                    Fig. \ref{H2_HW2} & 1.98170 & 0.04450 & 0.01305\\
                    Fig. \ref{H2_HW3} & 1.97752 & 0.09600 & 0.05558 \\
                    Fig. \ref{H2_HW4} & 1.97282 & 0.33308 & 0.03247\\
                    Fig. \ref{H2_HW5} & 1.96710 & 0.16029 & 0.05677\\
                \hline\hline
            \end{tabular}%
            \caption{Errors in the real part of Green's function with the  deviation in the number of electrons for multiple hardware runs for the H$_2$ molecule. The column labeled max error represents the maximum error that was observed in the G$_{00}$ element at any frequency in the grid. The deviation can also be seen when comparing to actual plots that are listed in  the Figure column in the table.}
            \label{hw_results}
        \end{table}
\subsection{Simulations on quantum hardware}
\label{hardware}

\begin{figure}[!ht]
\centering
\includegraphics[width=\columnwidth]{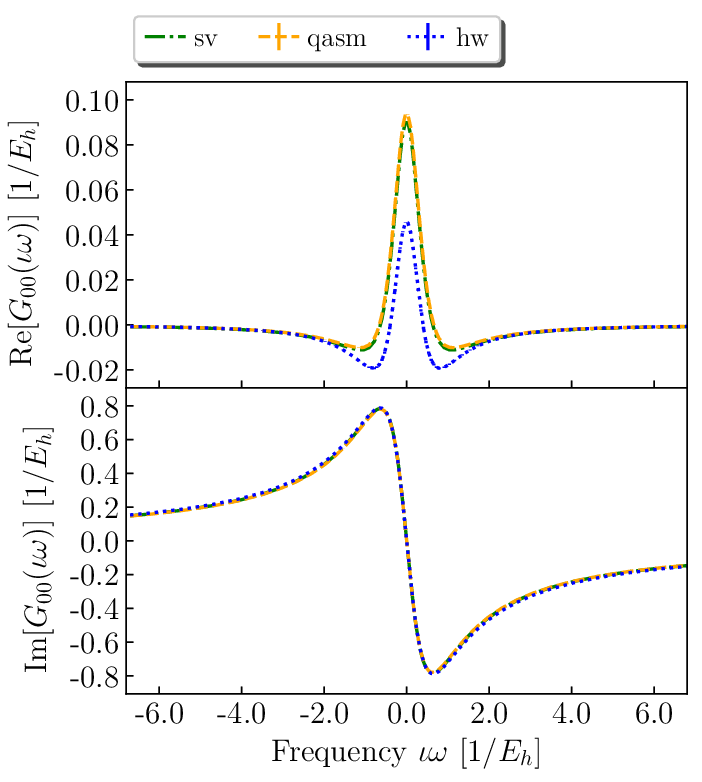}
\caption{Real (top) and imaginary (bottom) part of $G_{00}(\iota\omega)$ for the H$_2$ molecule with a bond length of 0.76 {\AA} in the STO-6G basis evaluated using QSE on $\mathsf{statevector}$ simulator, QASM simulator and hardware. The quantum hardware results have been post-processed using jackknife for a proper error propagation analysis. The vertical bars represent the error-bars.}
\label{H2_HW2}
\end{figure}

\begin{figure}[!ht]
\centering
\includegraphics[width=\columnwidth]{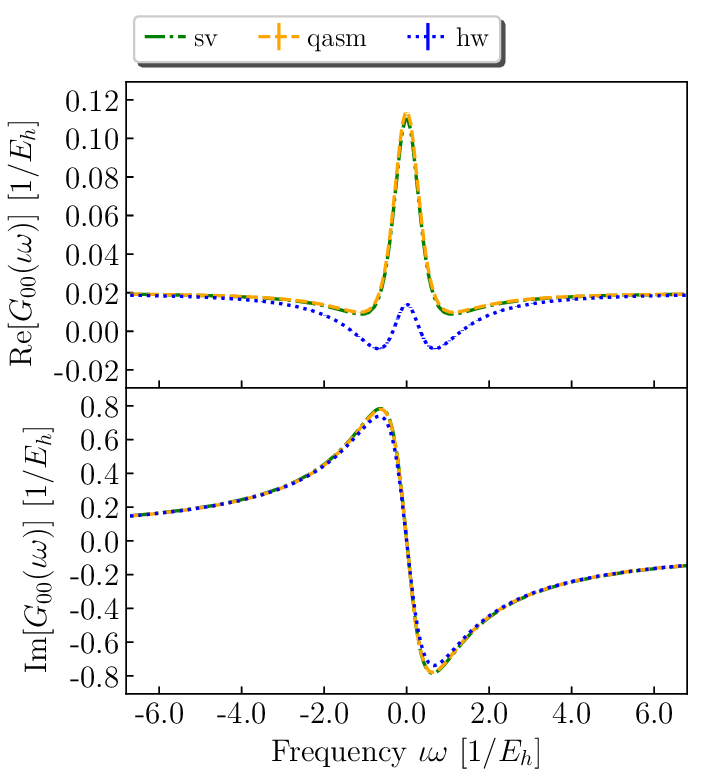}
\caption{Real (top) and imaginary (bottom) part of $G_{00}(\iota\omega)$ for the H$_2$ molecule with a bond length of 0.76 {\AA} in the STO-6G basis evaluated using QSE on $\mathsf{statevector}$ simulator, QASM simulator and hardware. The quantum hardware results have been post-processed using jackknife for a proper error propagation analysis. The vertical bars represent the error-bars.}
\label{H2_HW3}
\end{figure}

\begin{figure}[!ht]
\centering
\includegraphics[width=\columnwidth]{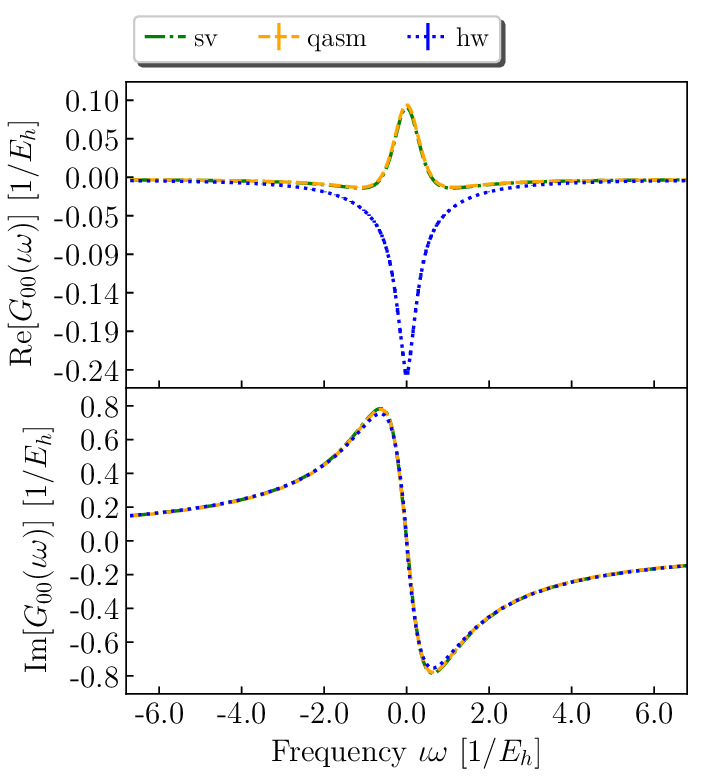}
\caption{Real (top) and imaginary (bottom) part of $G_{00}(\iota\omega)$ for the H$_2$ molecule with a bond length of 0.76 {\AA} in the STO-6G basis evaluated using QSE on $\mathsf{statevector}$ simulator, QASM simulator and hardware. The quantum hardware results have been post-processed using jackknife for a proper error propagation analysis. The vertical bars represent the error-bars.}
\label{H2_HW4}
\end{figure}

\begin{figure}[H]
\centering
\includegraphics[width=\columnwidth]{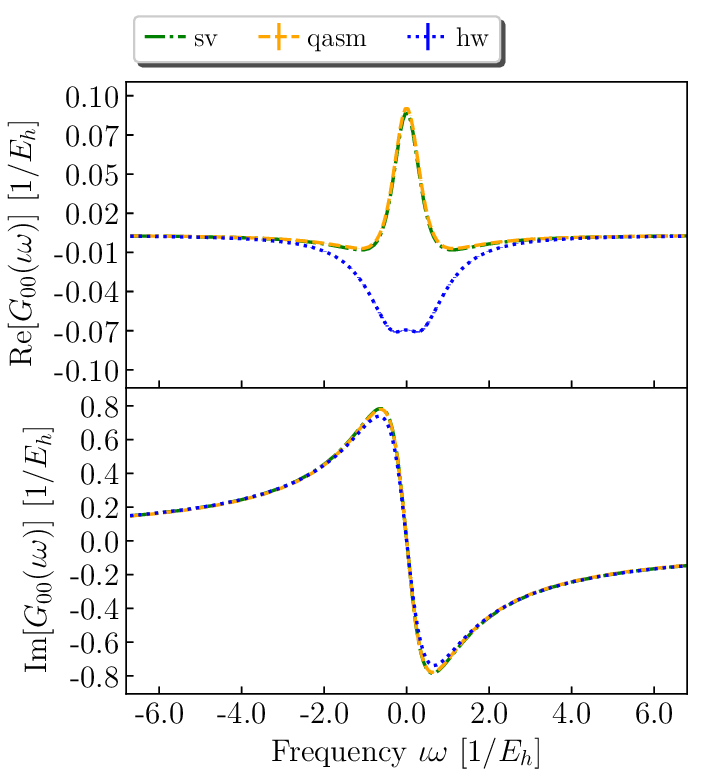}
\caption{Real (top) and imaginary (bottom) part of $G_{00}(\iota\omega)$ for the H$_2$ molecule with a bond length of 0.76 {\AA} in the STO-6G basis evaluated using QSE on $\mathsf{statevector}$ simulator, QASM simulator and hardware. The quantum hardware results have been post-processed using jackknife for a proper error propagation analysis. The vertical bars represent the error-bars.}
\label{H2_HW5}
\end{figure}

Finally, we want to understand how the results behave when the calculations are performed on the noisy quantum devices that exist today. This can be done in two different ways, either by using a simulator with a noise model or by running the calculations on an actual quantum machine itself. In this section, we show the results from the IBM hardware, $\mathsf{ibm\_hanoi}$. Figures ~\ref{H2_HW}-\ref{H2_HW5} show the $G_{00}(\iota\omega)$ element of the Green's function for H$_2$ molecule at the equilibrium geometry obtained through multiple hardware runs. We use read-out error mitigation to improve the results from the hardware, and re-sample them using the jackknife technique to ensure a correct error-propagation. Through multiple hardware runs, we note that the real part of the Green's function is more sensitive to quantum errors. We observe that the deviation in the real part of Green's function is directly proportional to the deviation in the number of electrons. As can be seen from Table~\ref{hw_results}, a larger deviation in the number of electrons points towards a higher error in the Green's function elements.
Note that such a dependence of the error in the number of electrons and the real part of the Green's function is completely expected. In the Green's function theory, the number of electrons in the system can be evaluated from the the real part of the Matsubara Green's function in the frequency domain.

\section{Conclusions}
In this work, we described a hybrid quantum-classical approach for calculating Green's functions on quantum computers. We start by approximating the ground state of the system using VQE. We expand the subspace around this VQE ground state to obtain excited-state energies and transition matrix elements for $N+1$ and $N-1$ states, where $N$ is the number of electrons in the ground state. These transition matrix elements are then used on a classical machine to calculate the Matsubara Green's function in the Lehmann's representation.

We observed that, in the absence of quantum errors and for the systems considered in this work, QSE produces the Green's function in reasonable agreement with the FCI Green's function. In more complex situations, the QSE subspace chosen here may lead to inaccurate results, and a larger subspace may be required.
We further showed how this algorithm behaves in the presence of quantum measurements. This was followed by a discussion about the propagation of statistical noise through non-linear mathematical operations using re-sampling techniques.

Finally, results from quantum hardware were included, to demonstrate how the algorithm behaves in presence of decoherence affecting the near-term devices. We employ multiple error-mitigation techniques to minimize the noise from the hardware. We observe that hardware results are not always exact, however, we are able to reproduce the Green's function with significant accuracy and satisfactory error bars. Here, we show hardware results for $H_2$ molecule. In the future work, this procedure can be extended to bigger systems. The number of matrix elements needed will increase when scaling up to larger systems, but these calculations can be done in parallel over multiple processors. \\

We believe that the ability to calculate Green's functions on near-term quantum computers is an opportunity to study chemical systems in more detail. In particular, Green's function based approaches allow us to calculate various thermodynamic and excited-state properties. The algorithm proposed here can be used in the framework of embedding theories such as DMFT and SEET. Furthermore, the approach proposed here can also be used in conjunction with dynamic self-energy mapping (DSEM) to calculate the Green's functions on near-term quantum devices.\\
\section*{Acknowledgements}
D.Z. and D.D.
were supported by the
“Embedding QC into Many-Body Frameworks for Strongly
Correlated Molecular and Materials Systems” project, which
is funded by the U.S. Department of Energy, Office of
Science, Office of Basic Energy Sciences, Division of
Chemical Sciences, Geosciences, and Biosciences. 

\bibliographystyle{apsrev4-1}
\bibliography{main}
\end{document}